\documentclass[twocolumn,pre,showpacs]{revtex4-1}
\usepackage{amssymb,latexsym,mathrsfs,amsfonts,amsmath,graphicx,color,wrapfig, hyperref,color}

\begin{document}
 
\title{Dependence of asymptotic decay exponents on initial condition and the resulting scaling violation}

\author{Sourish Bondyopadhyay}
\email{sourish.bondyopadhyay@gmail.com}
\affiliation{CMP Division, Saha Institute of Nuclear Physics,
1/AF Bidhan Nagar, Kolkata 700064, India.}

 
\begin{abstract}
There are several examples which show that the critical exponents can be dependent on initial condition of the system. In such situations, there are many systems where various issues related to the universal behavior {\it e.g.} existence of universality, splitting of universality class, scaling violation, whether the initial dependence should persist even after sufficiently long time or is a transient effect, the reasons for such features, etc. are not yet quite clear. In this article, with the simple example of conserved lattice gas  (CLG) model, we investigate such issues and clearly show that under certain situations the asymptotic decay exponents are in fact dependent on the initial condition of the system. We show that such effect arises because of existence of two competing time scales, and identify the initial conditions which capture the universal features of the system.

\end{abstract}

 
\maketitle 
 
 
\section{Introduction:} 
Study of universality class is of fundamental interest in order to understand the basic principles of continuous phase transitions \cite{Henkel,Marro,Lubeck}. A paradigmatic example of universality class for non-equilibrium systems, is directed percolation (DP) \cite{dp}. Some of the best studied examples of non-equilibrium systems are contact process \cite{cp}, epidemic spreading \cite{epd}, reaction diffusion process \cite{Odor}, various sandpile models like Manna model \cite{Manna}, conserved lattice gas (CLG) \cite{Rossi,CLG_Oliveira,CLG_Lee}, conserved threshold transfer process (CTTP) \cite{Rossi,CTTP_contracdictory}, Maslov-Zhang sandpile \cite{Maslov}, etc. Most of the studies in this direction are numerical. 

Determination of universality class strongly relies on the determination critical exponents. There are several examples where it is found that the critical exponents can depend on the initial condition of the system. Examples include interface growth models, annihilating random walk (ARW), branching and annihilating random walk (BARW),  production and annihilating random walk (PARW) {\it etc.} and their several variants \cite{Odor}. It is usually believed that special symmetries, additional conservation, presence of different sectors {\it etc.} are the key ingredients for the initial condition dependence. However, there are several examples which show that initial dependence arises even in absence of these things, claiming that those reasons are irrelevant \cite{Odor}. Even in systems where special symmetries, additional conservation {\it etc.} may be relevant to the universality class of the system, their relevance to the initial condition dependence are not guaranteed. Besides, there are systems where the 
spreading exponents {\it e.g.} growth exponents or the critical initial slip exponents, survival probability exponents {\it etc.} were found to depend on  initial condition \cite{Jensen, Init}. But later it was found \cite{Tran} that in case of spreading exponents, such features might be transient (short-time) behavior of the system and eventually should  cross over to the usual universal decay after sufficiently large time. Thus, there are many systems where the initial  dependence and the reasons are still debated.

Now, universality class means that different systems undergoing phase transition are characterized by the same kind of critical behavior and hence the same set of critical exponents, irrespective of the microscopic details. Assume that for a particular system different initial conditions lead to different values of the dynamic exponents.  So, following are some questions which may immediate arise. Are  what is universal here ? Are the critical behavior of the system non-universal ? Does the universality class split into a number of subclasses, with each initial condition having its own subclass ? On the other hand if it is believed that there should be an unique universality class for the system, then which initial condition should be relied and why not the others ? Should the scaling relations be violated ? Should the critical behavior for  the different initial conditions converge to an unique behavior after sufficiently large time ? For various systems, there are several issues where  understanding of 
initial condition dependence of the critical behavior, is not quite clear. In this article we try to clarify this kind of issues.  

In this article we have studied a simple model, Conserved Lattice Gas (CLG) in 1D. For this model it has been claimed earlier \cite{CLG_Lee} that the asymptotic decay exponent is $\alpha=1/4$ and the scaling relation $\alpha=\beta/{\nu_\parallel}$ is satisfied, but $z=\nu_\parallel/\nu_\perp$ is violated. Here we clearly show that $\alpha$ depends on the choice of initial condition (i.c.): $\alpha_{in}=1/4$ in case of random i.c., and $\alpha_{in}=1/2$ in case of natural i.c. which is in agreement with that obtained from the decay of autocorrelation with time ($\alpha_{ss}$) measured in the stationary state. The subscript $in$ indicates that $\alpha$ depends on initial condition and the subscript $ss$ stands for stationary state measure of $\alpha$. The exponent consistent with the scaling relations is $\alpha_{in}=\alpha_{ss}=1/2$. We show that such situation happens due to existence of two  competing time scales.

\newpage
\section{Conserved Lattice Gas (CLG):}
The model \cite{Rossi}  was introduced  as an exclusion process with nearest neighbor repulsion. Each site is either occupied by a single particle or empty. The dynamics is such that a site is active if it is occupied by a particle and has at least one neighboring site which is occupied, and at least one vacant neighbor. The particle from the active site hops to one of the neighboring vacant sites. In 1D the active site has one occupied neighbor and the other neighbor is empty. It is a closed system with periodic boundary condition. The density of particles $\rho$, remains conserved. Total number of active sites $(\langle 110\rangle+\langle011\rangle)$, is considered as the order parameter. Under the tuning of $\rho$, the model undergoes a continuous transition from an active phase to an absorbing state, at $\rho_c=1/2$. Simplicity of the model makes it particularly interesting in order to get some insight into the basic principles of phase transition and universality.

 
In 1D the stationary state of the model  is exactly solvable \cite{CLG_Oliveira}. The static exponents ({\it i.e.} exponents dependent on the stationary state only)  $\beta$ and $\nu_\perp$ are known exactly. However, the  dynamic exponents ({\it i.e.} exponents dependent on the time evolution of the system) $\alpha,~\nu_\parallel,$ and $z$ were determined from numerical simulations. The symbols used, are standard conventional symbols for absorbing phase transition \cite{Henkel,Lubeck}.
The order parameter which is a function of time $t$, distance from the criticality $\Delta$ and system size $L$, can be expressed as $\rho_a(t,\Delta,L)=t^{-\alpha} F(1, t^{1/\nu_\parallel}\Delta,t^{-1/z}L)$, where $F$ is the scaling function. Under extreme cases of the  variables, we get the following equations:
$\rho_a(t,\Delta,\infty)=t^{-\alpha} f(t \Delta^{\nu_\parallel}); ~\rho_a(t,0,\infty)\sim t^{-\alpha};~
 \rho_a(\infty,\Delta,\infty)\sim\Delta^\beta;~~\rho_a(t,0,L)=t^{-\alpha} g(t/L^z);~
 \rho_a(\infty,\Delta,L)=L^{-\beta/\nu_\perp} h(L\Delta^{\nu_\perp})$ where, $f,~g,~h$ denote the scaling functions. The above expressions lead to the scaling relations
$\alpha=\beta/\nu_\parallel,~ z=\nu_\parallel/\nu_\perp$. 
For determination of $\nu_\parallel$ from the subcritical data we use the relation 
$\rho_a(t,\Delta,\infty)=t^{-\alpha} f(t \Delta^{\nu_\parallel})$, using the estimate for $\alpha$. To obtain  $\nu_\parallel$  from the supercritical data, we use the convenient form $\rho_a(t,\Delta,\infty)=\rho_a  k(t\Delta^{\nu_\parallel})$ which do not require estimate for $\alpha$,  where $\rho_a $ is the stationary value of $\rho_a(t)$, and $k$ is the scaling function.
The existing results \cite{CLG_Lee} for the static and dynamic critical exponents of 1D CLG are listed in table \ref{tab}.


\section{Scaling Violation and Its Resolution:}
From the  list of exponents, one important thing to notice is that for the existing results $\alpha=\beta/\nu_\parallel~ ,~~ \mbox{but}~~ z\ne \nu_\parallel/\nu_\perp$.
This violation of scaling has been reported
\begin{table}[h!]
\begin{center}
\begin{tabular}{|c|c|c|c|c|c|c|c|}\hline
\multicolumn{2}{|c|}{Exact results \cite{CLG_Oliveira}} & \multicolumn{3}{|c|}{Existing results \cite{CLG_Lee}} & \multicolumn{3}{|c|}{Our results} \\ \cline{1-8}
~~~~$\beta$~~~~ & ~~~~$\nu_\perp$~~~~&~~~$\alpha$~~~ &~~~$\nu_\parallel$~~~&~~~ $z$~~~&~~~$\alpha$~~~ &~~~$\nu_\parallel$~~~&~~~$z$~~~\\ \hline
1 & 1 & 1/4 & 4 & 2 & 1/2 & 2 & 2\\ \hline
\end{tabular}
\end{center}
\vspace{-0.4 cm}
\caption{Our new estimates for the critical exponents of 1D CLG are compared with the existing results.}
\label{tab}
\end{table}
\begin{figure}[!h]
\vspace{-0.2 cm}
\includegraphics[width=4.4 cm, height=4 cm]{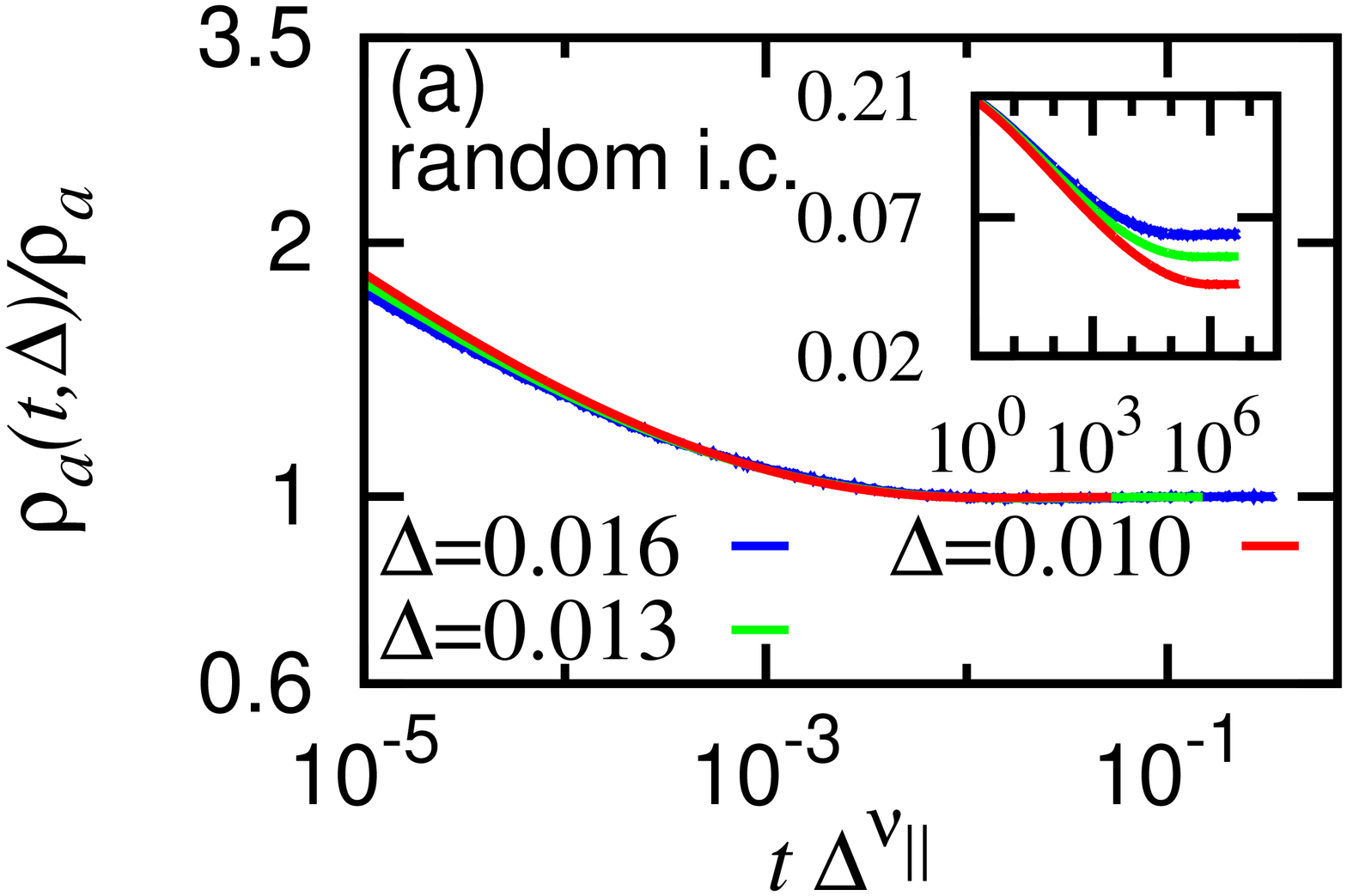}\includegraphics[width=4.4 cm, height=4 cm]{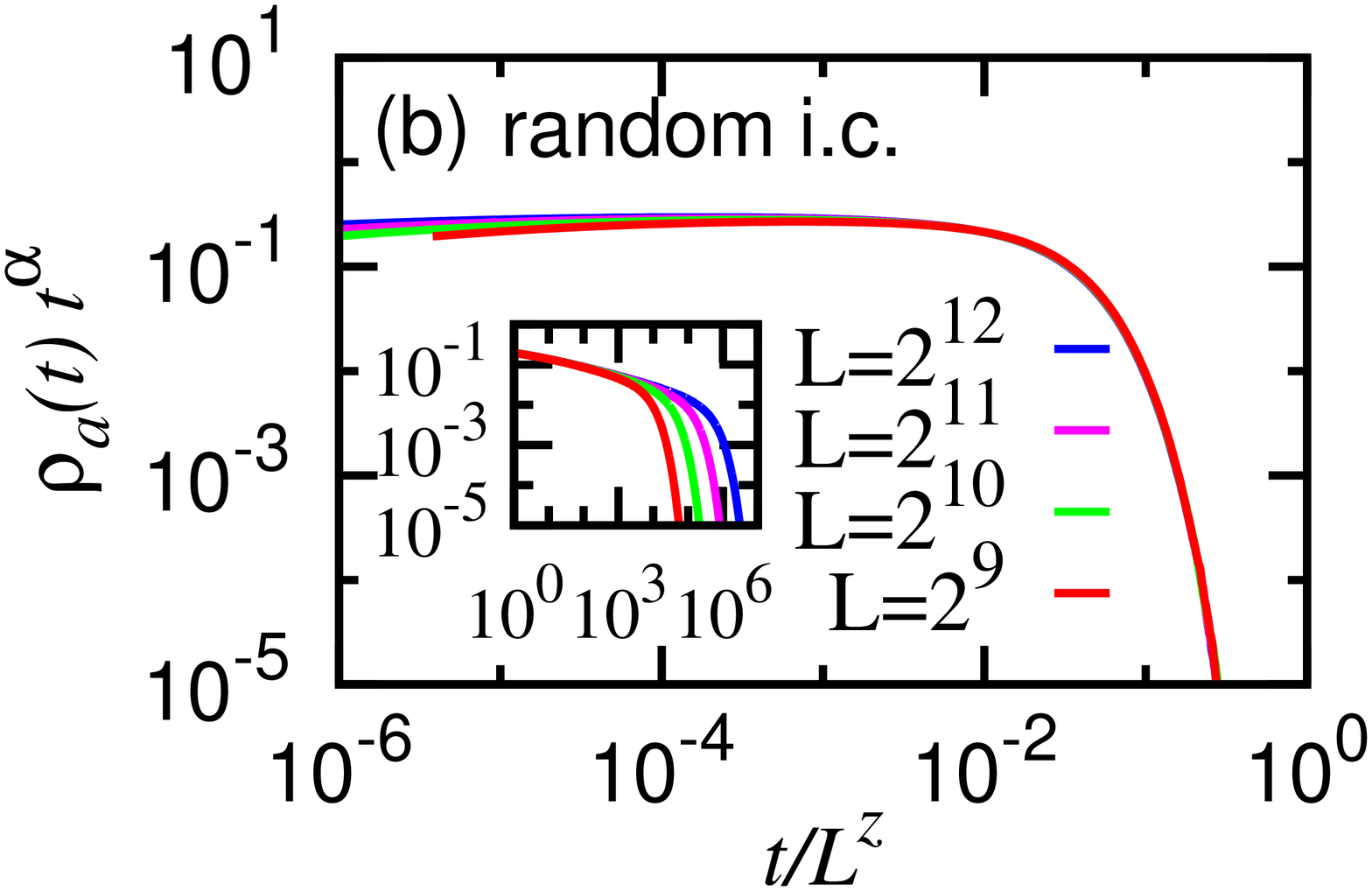}
\vspace{-1 cm}
\caption{(Color online) Using random i.c. determination of (a) $\nu_\parallel$ from supercritical data; (b) $z$ from data at criticality; Insets show the unscaled data.}
\vspace{-0.3 cm}
\label{fig1}
\end{figure}
earlier \cite{CLG_Lee}. There are also several other examples of scaling violation \cite{Rossi, Lubeck_split} in the literature, particularly in one of the two scaling relations mentioned above.   

\begin{figure}[h!]
\includegraphics[width=4.4 cm, height=4 cm]{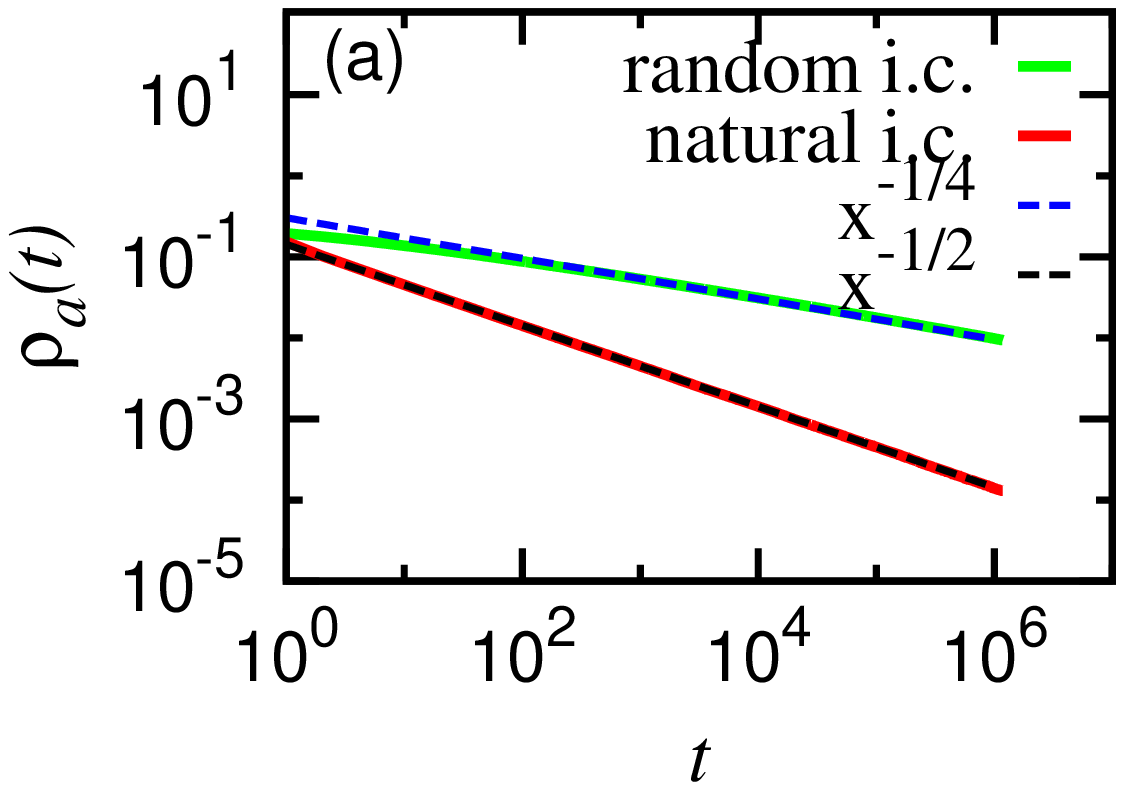}\includegraphics[width=4.4 cm, height=4 cm]{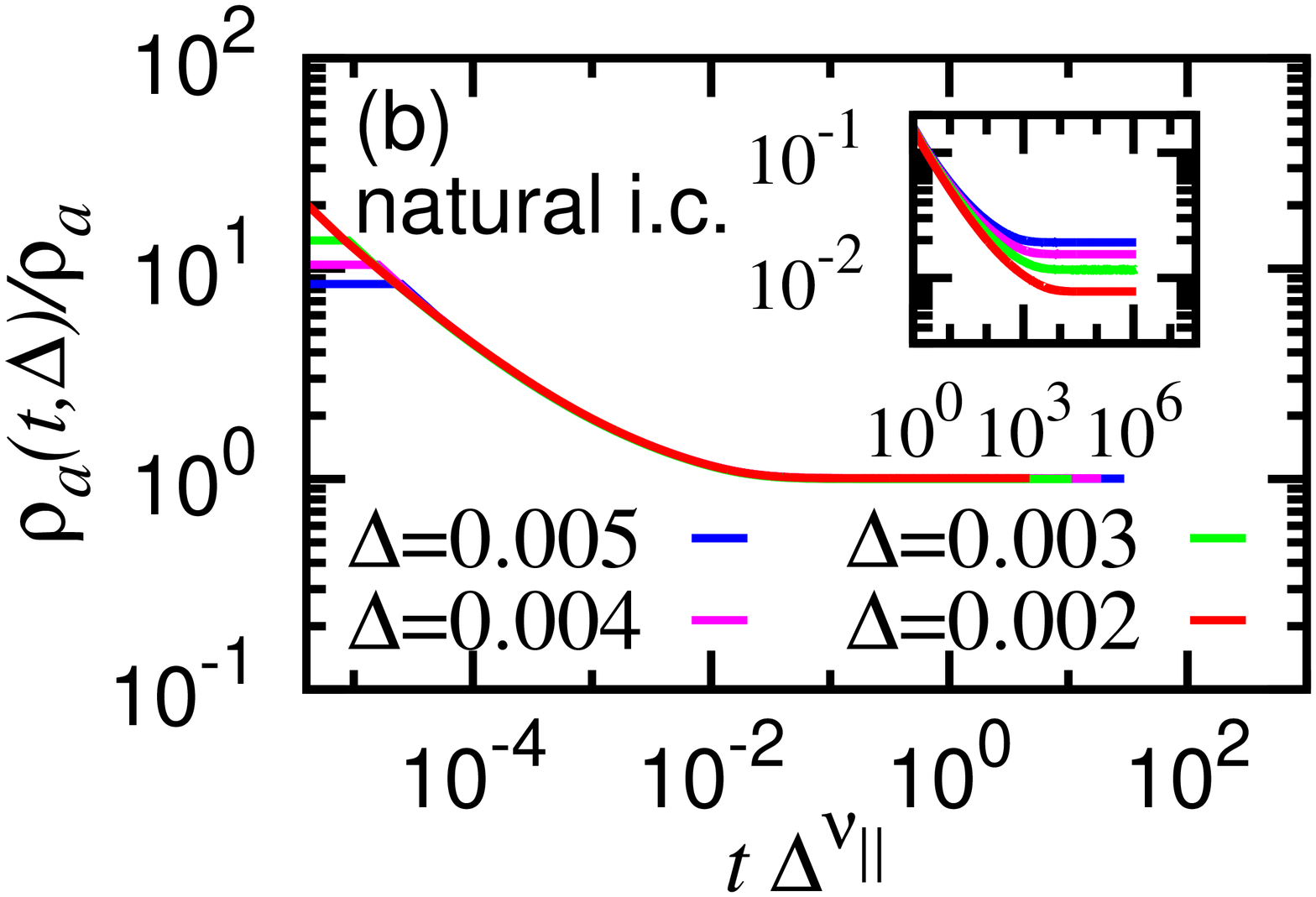}
\includegraphics[width=4.4 cm, height=4 cm]{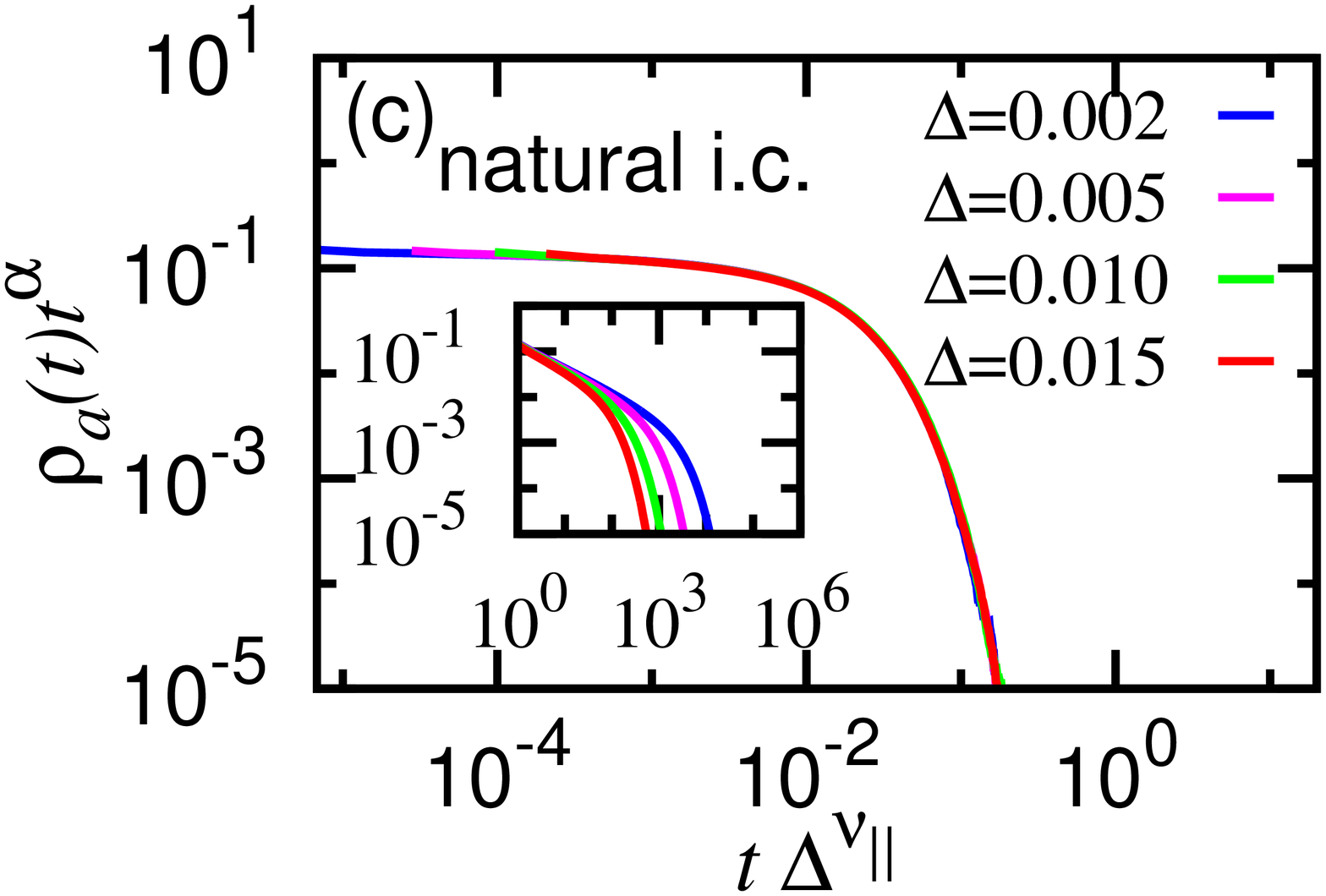}\includegraphics[width=4.4 cm, height=4 cm]{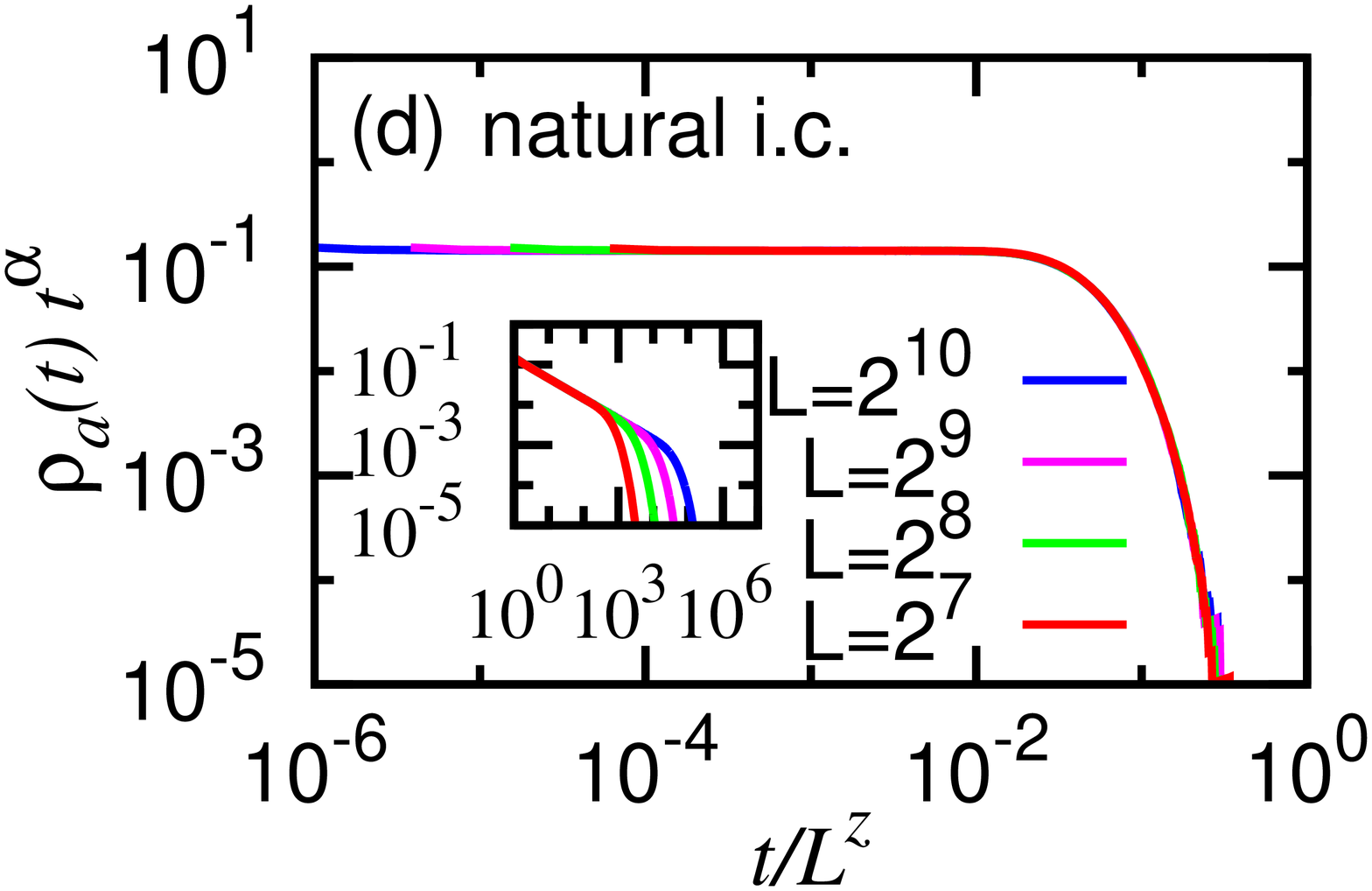}
\vspace{-0.9 cm}
\caption{(Color online) (a) Decay behavior for random i.c. and natural i.c.; Using natural i.c. data collapse for the determination of (b) $\nu_\parallel$  from  supercritical data;  (c) $\nu_\parallel$ from subcritical data; (d) $z$ from the data at criticality; Insets: Unscaled data.}
 \vspace{-0.2 cm}
 \label{f2}
\end{figure}

Here, we have used natural initial condition \cite{Wilczek,Jensen,fes_dp} instead of random initial condition. The exponents obtained from natural i.c., listed in table \ref {tab}, are in agreement with those obtained from stationary state autocorrelation, resulting in the set of exponents consistent with the scaling relations. Thus, the scaling violation in 1D CLG is resolved. The simulation results for random and natural i.c. are shown in Figs. \ref{fig1}
and \ref{f2} respectively. The decay behaviors for these two i.c.'s are
compared in Fig. \ref{f2}(a).

In the following paragraph we explain the natural i.c.

\vspace{-0.5 cm}
\section{Natural initial condition:}
\begin{figure}[!h]
\includegraphics[width=8.5 cm, height=4 cm]{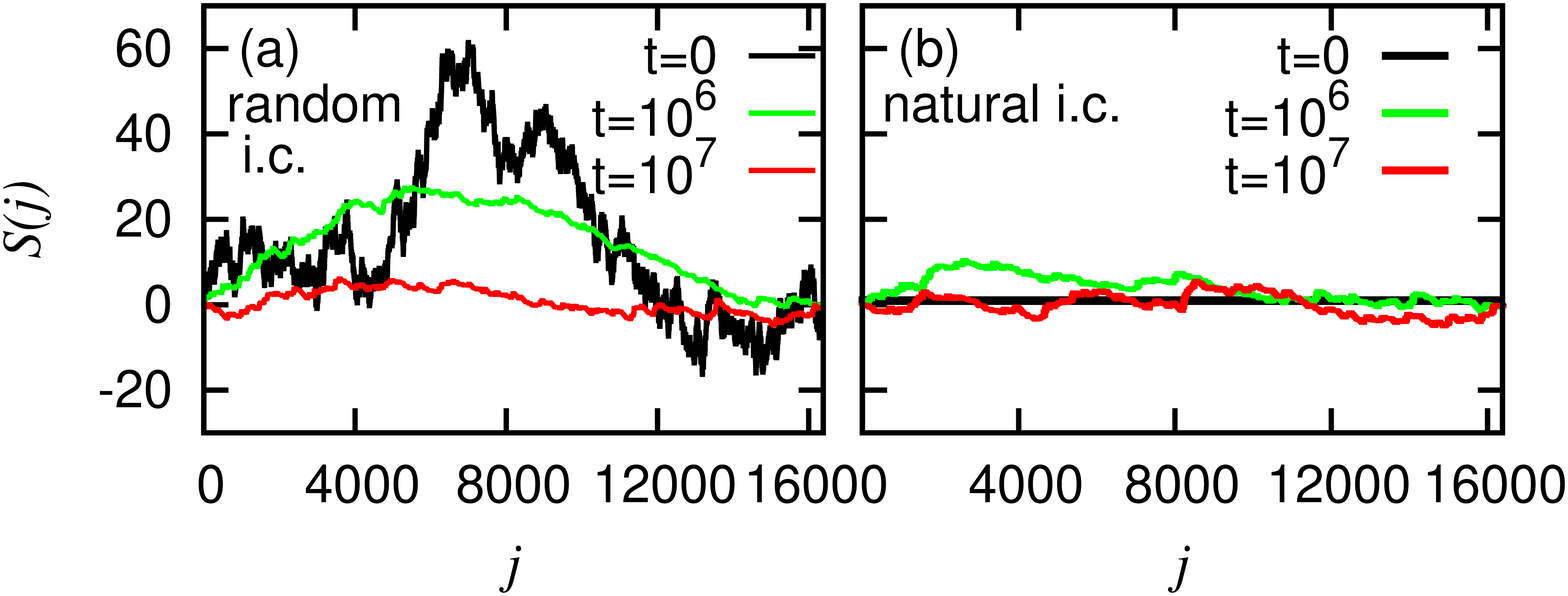}
\vspace{-0.3 cm}
\caption{(Color online) (a) Cumulative background density $S(j)$ for random i.c. captured at different time. (b) The same for natural i.c.}
\vspace{-0.0 cm}
\label{fig2}
\end{figure}

Here we use a correlated uniform initial condition following a work  by  Toussaint and Wilczek \cite{Wilczek}, which has been used later in many other articles \cite{Jensen,fes_dp}. It is a special  initial condition which carries the long range correlations of the stationary state of the system, from the beginning, so that the system is spatially uniform (density profile homogeneous) at the very beginning and  it  remains uniform throughout the evolution of the system with time. In this sense uncorrelated random i.c. {\it i.e.} filling the sites with particles randomly, is not uniform or homogeneous. The density profiles for random i.c. and natural i.c. are shown in Fig. \ref{fig2} where the cumulative sum $S(j)=\sum_{i=1}^j n_i-j N/L$  is a measure of excess particles in the spatial region from the $1^{st}$ site to $j^{th}$ site, with respect to the expected average, and $n_i$ denotes the number of particles at site $i$, $N$ is the total number of particles, $L$ is the system size.
Even though the density 
fluctuation for both random i.c. and natural i.c., is proportional to $\sqrt L$, but the constant of proportionality is very small in case of natural i.c., compared to that in random i.c. and hence the effect of density fluctuation is negligible in natural i.c. Although perfect uniform or homogeneous density profile is not possible when the system is evolving with time, Fig. \ref{fig2}  shows that for practical purposes natural i.c. can be considered as uniform or homogeneous. With random i.c., it takes a long time for the system to reach a uniform state. 

To produce  random i.c., following the usual convention we start with an empty lattice, choose the sites randomly and fill the vacant sites until the total number of particles reach the desired value $\rho L$. Now, as described in this section, natural i.c. is a suitably reactivated stationary state of the system, where the reactivation is needed to start with a highly active state from which the decay behavior can be studied for some considerable decades of time. Depending on the dynamics, a suitable reactivation can work better since the stationary state is perturbed while reactivating. Here, in 1D CLG, at the critical point $\rho_c = 1/2$, the only stationary state is of the form \{\ldots 101010 \ldots\}. For $\rho>1/2$, the excess particles take the places of the zeros in the previous configuration in a spatially homogeneous manner. Since we know the stationary states, we perform the process of preparing natural i.c.  by producing a spatially homogeneous reactivated stationary state of the form \{\ldots 
11001100 \ldots\} first and then put (remove) excess particles at vacant (occupied) sites in almost equal spatial interval to obtain the homogeneous state with the desired density $\rho$. Natural i.c. can also be produced in a way as suggested in the article \cite{fes_dp}. In that case the system starts from a random i.c., and is evolved until it reaches a stationary state. Then it is reactivated by diffusion for one Monte Carlo cycle, and  that can serve as the initial condition. It is important to note that natural i.c. is not just a flat initial condition.

Thus we find that the evolution of the system with time and hence the decay exponent is not unique, it is rather dependent on the initial condition of the system. The natural i.c. gives drastically different result compared to the usual random i.c. and the exponents resulting from natural i.c. are consistent with the scaling relations whereas the exponents obtained from random i.c., show scaling violation.

\section{Decay exponent from stationary state autocorrelation:}
The autocorrelation function is defined as $c(\delta t)=\langle \chi_{i}(t)\chi_{i}(t+\delta t)\rangle-\langle \chi_{i}(t)\rangle\langle \chi_{i}(t+\delta t)\rangle $ where $\chi_{i}(t)$ is the site variable for activity at time $t$ such that $\chi_{i}(t)=0$ if site $i$ is inactive and equal to 1 otherwise. Since, in the stationary state $\langle \chi_{i}(t)\rangle=\langle \chi_{i}(t+\delta t)\rangle$, the stationary state autocorrelation function is given by $c_{ss}(\delta t)=\langle \chi_{i}(t)\chi_{i}(t+\delta t)\rangle-\langle \chi_{i}(t)\rangle^2$, where the subscript $ss$ denotes that $c_{ss}(\delta t)$ is a stationary state measurement.

Now, the function  $\langle \chi_{i}(t)\chi_{i}(t+\delta t)\rangle$ can be expressed as $\langle \chi_{i}(t)\chi_{i}(t+\delta t)\rangle=(\delta t)^{-\alpha_{ss}} f(\delta t \Delta^{\nu_\parallel})$, where $\Delta=(\rho-\rho_c)$ is the distance from the criticality, $f$ is the scaling function, and $\nu_\parallel$ is a critical exponent, explained in earlier sections. In the critical regime the system remains correlated up to time  $\xi_\parallel\sim \Delta^{-\nu_\parallel}$. Thus, for $\delta t<\xi_\parallel$ {\it i.e.} before reaching the steady value $\rho_a^2$,  it is expected that $\langle \chi_{i}(t)\chi_{i}(t+\delta t)\rangle$  scales in the same way as that of $\rho_a(t)$ and hence $\alpha_{ss}$ should be same  as the order parameter decay exponent {\it i.e.} $\alpha_{ss}=\alpha$ (see P 106-107 in \cite{Henkel}). Thus one can obtain the decay exponent $\alpha$ that describes the time evolution of the order parameter at the critical point, from the decay of autocorrelation in the stationary state.
 
\begin{wrapfigure}{r}{0.23\textwidth}
\vspace{-0.7 cm}
 \hspace{-0.4 cm}
\includegraphics[width=4.4 cm, height=4 cm]{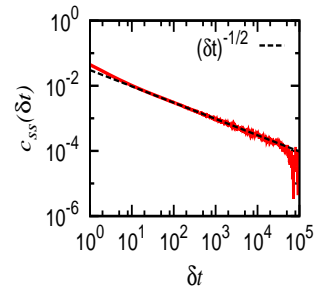}
\vspace{-0.7 cm}
\caption{(Color online) Decay of autocorrelation in stationary state.}
\vspace{-0.4 cm}
\label{f4}
\end{wrapfigure}
In order to obtain $\alpha_{ss}$, one can start with any initial condition in the supercritical regime ($\rho>\rho_c$) and let the system evolve until it reaches the stationary state. Then the autocorrelation $c_{ss}(\delta t)$ is measured. Here, we obtain $c_{ss}(\delta t)\sim (\delta t)^{-1/2}$ {\it i.e.} $\alpha_{ss}=1/2$, as shown in Fig. \ref{f4}. Since the stationary state is unique, it has no memory of the initial condition. Hence it is expected to be free from any anomalous behavior which could arise from different type of initial conditions. However, it is customary to measure $\alpha$ from the decay of order parameter at criticality since in the stationary state within the critical regime the saturation value of the order parameter itself is usually very small which results in  smaller value of $c_{ss}(\delta t)$. This makes it difficult to measure $\alpha$ from the decay of autocorrelation. But, unless it is too small to measure, the decay behavior of autocorrelation should be free from anomalous 
features and captures the universal decay feature.

\section{Analytical Arguments:}
So far we have talked about numerical simulation of CLG. 
Now, let us consider the analytical arguments for the time evolution of the system. 

In fact the spreading exponents for the natural i.c. can be easily determined analytically and  one can obtain the decay exponent $\alpha$ from the spreading exponents assuming that the hyperscaling relations hold. It is customary to obtain the spreading exponents from the following relations- $P_{sur}(t)\sim t^{-\delta},~N_a(t)\sim t^{\Theta}$, where $P_{sur}(t)$ is the survival probability, $N_a(t)$ is the number of active sites, and $\delta,~\Theta$ are the survival probability exponent and the slip exponent respectively. The exponents satisfy the hyperscaling relation  $z(\Theta+\delta)=d-\beta/\nu_\perp$ (see P 268 in \cite{Henkel}), where $d$ is the system dimension  and $\beta,~\nu_\perp,~z$ are critical exponents explained in earlier sections. Knowing the spreading exponents, one can obtain the decay exponent $\alpha$ using the hyperscaling relation $\Theta=d/z-\alpha-\delta$  (see P 268 in \cite{Henkel}).

For determination of spreading exponents first we consider the stationary state at the critical point which is of the form $\{\ldots 101010 \ldots\}$. The system is perturbed by moving a particle to one of its vacant neighbors so that a $11$ pair and a $00$ pair are formed there by producing a pair of active sites \cite{CLG_Oliveira}. The time evolution of the system can be viewed as just the diffusion of a single particle in one dimension in presence of an absorbing boundary where the $00$ pair plays the role of the absorbing boundary. The random walk continues until the system reaches the absorbing state. In such case it is well known that the survival probability varies with time as $P_{sur}(t)\sim t^{-1/2}$. Knowing the exponents $\beta=1,~\nu_\perp=1,~z=2,~\delta=1/2$ the hyperscaling relation results in $\Theta=-1/2$. With this one can easily obtain $\alpha$ using the other hyperscaling relation, which gives $\alpha=1/2$.

Moreover, the dynamics of the model is very similar to the model of  particle-antiparticle annihilation in diffusive motion, which is very well known \cite{Wilczek,Redner}.  In this context the article \cite{Wilczek} by Toussaint and Wilczek, is particularly interesting from our point of view. In this model there are two species- particles ($A$) and antiparticles ($B$). They move diffusively ($A0\rightleftarrows 0A,~B0\rightleftarrows 0B$) and annihilate ($AB\to 00,~BA\to 00$) each other  whenever they come together.  It was analytically shown that for random initial condition $\rho_a(t)\sim t^{-1/4}$ whereas for a correlated uniform initial condition $\rho_a(t)\sim t^{-1/2}$. Same results hold when one of the species is stationary. It was  explained that in case of random initial condition regions containing mostly particles were separated from the regions which contained mostly the antiparticles and that retarded the entire process.

Now, consider the CLG dynamics- $1100\to 1010$ and $1101\to 1011$, and the symmetric counter parts, where $1$ and $0$ stand for particle and vacancy respectively. The sites are active only in presence of $11$ pairs. Thus, number of $11$ pairs {\it i.e.} $<11>$ is an equivalent order parameter. When ever a $11$ pair and a $00$ pair come together, both pairs are  annihilated (transformed to $10$ or $01$). Thus, the two processes $1100\to 1010$ and $1101\to 1011$, can be considered as annihilation and diffusion of $11$ pairs respectively. In the case of diffusion-annihilation model the particle-antiparticle pairs annihilate ($AB\to 00$) whenever they come side by side, and diffuse when they face vacancy ($A0\rightleftarrows 0A$). Thus, CLG and diffusion-annihilation model effectively share the same basic mechanism. Their decay behaviors are found to be the same as expected.
  
\section{Role of non-ergodicity and conservation:}
Now, for better understanding of the underlying mechanism, we focus on the simpler example of diffusion-annihilation process. In this case there is no scope for creation of particle or antiparticle and the system is non-ergodic. The unusual decay behavior may seem to be originated from the non-ergodicity. But, though these system undergoes absorbing transition, minor modifications ({\it e.g.} allowing $00\to AB$ with some probability $p$) will make it an equilibrium system and the limit $p\to 0$  also shows two distinct decay behavior depending on the initial condition. Thus, the non-equilibrium situation or ergodicity  has nothing to do with the unusual decay features. 

Parity conservation is another possibility. In terms of the annihilating random walk problem, since annihilation takes place in pairs ($AB\to 00$), parity is conserved globally. In case of the corresponding single component model the dynamics is $AA\to 00$, where parity is conserved locally. But, this model do not show any initial condition dependence which is contrary to the two component model. Thus, whether the parity conservation is relevant to the universality class or not, it seems to be irrelevant as far as initial dependence of critical behavior is concerned.

Another intuitive reason could be the  presence of an additional conserved quantity. Here, the difference between the number of particles and antiparticles $(A-B)$, is conserved since they are annihilated only in pairs $(AB\to 00)$. But, the model can be modified by an additional dynamics which allows $AA$ or $BB$ pairs to throw one from the pair to a randomly chosen vacant site in the system. This changes the decay behavior of the system such that there is an unique decay $\rho_a\sim t^{-1/2}$, even though the additional conservation is still present. In case of 1D CLG, the difference between the number of $11$ pair and $00$ pair $(\langle 11\rangle-\langle 00\rangle)$, is conserved since they are annihilated only in pairs $(1100\to 1010)$.
However, the two point spatial correlations alway satisfy the relations: $\langle 11\rangle+\langle 10\rangle=\rho$ and $\langle 00\rangle+\langle 01\rangle=1-\rho$ which hold irrespective of the dynamics. Thus, $\langle 11\rangle-\langle 00\rangle=2\rho-1$. Thus, the conservation automatically follows from the conservation of $\rho$. So, presence of the additional conservation may not be relevant to the unusual critical behavior.

\section{Plausible explanation:}
Here we propose and will show in the next section that such situation arises because of existence of two competing time scales. In general, the decay process of a system may show a initial behavior, and after that follows the asymptotic behavior which is usually unique for each system. Memory effect give rise to long range spatial and temporal correlation. Depending on the dynamics it takes some time for the system to grow the long range fluctuations and suppress microscopic details including the memory of the initial condition. What really matters is the initial time scale ($\tau_{in}$) corresponding to the initial memory and it depends on the spatial extent  $l_{is}$ of the largest island in the density profile (Fig. \ref{fig2}). The time scale $\tau_{in}$ corresponds to the time taken by the largest island to spread (diffuse) and reach a uniform state, and it is given by $\tau_{in}\sim l_{is}^2$. There is another time scale $\tau\sim L^z$, imposed by the finite size of the system. What actually happens 
in case of CLG with random i.c., is that these two time scales $\tau_{in}$ and $\tau$, are comparable to each other $(\sim L^2)$ and since  the initial process corresponds to $\rho_a(t)\sim t^{-1/4}$ which is slower, the universal decay behavior $\rho_a(t)\sim t^{-1/2}$, is completely suppressed. We will show in the following paragraphs how $\tau_{in}$ can be controlled.

\section{Controlling the initial time scale $\tau_{in}$:}
Here, we use some special initial conditions to show how the initial time scale $\tau_{in}$ can be adjusted properly. Consider a correlated blocked initial condition such that all the particles are placed in a single block of size $L/2$ in the form $\overbrace{\ldots 0000}\overbrace{1111\ldots}$ (periodic boundary condition is assumed).  
\begin{figure}[h!]
\includegraphics[width=4.4 cm, height=4 cm]{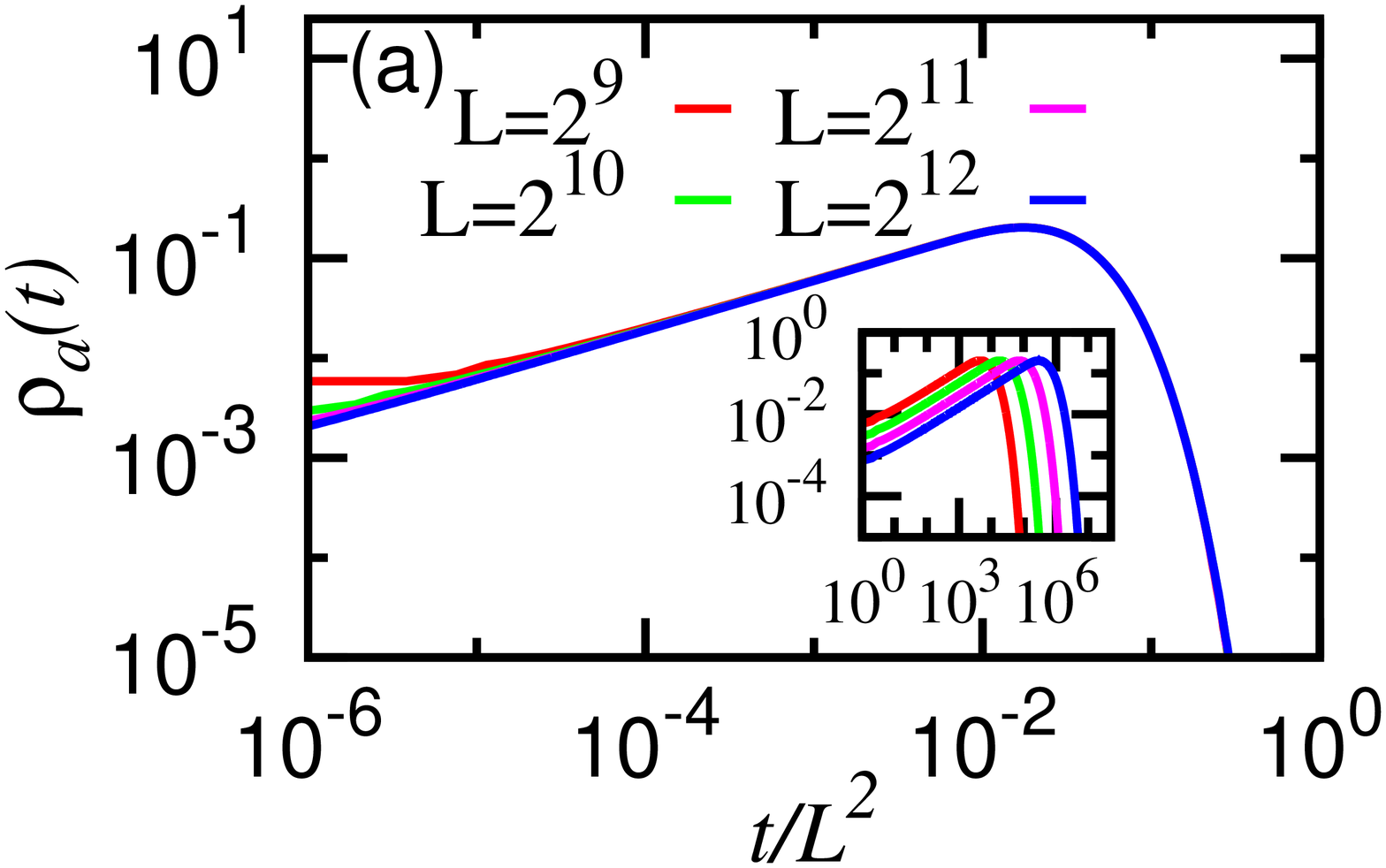}\includegraphics[width=4.4 cm, height=4 cm]{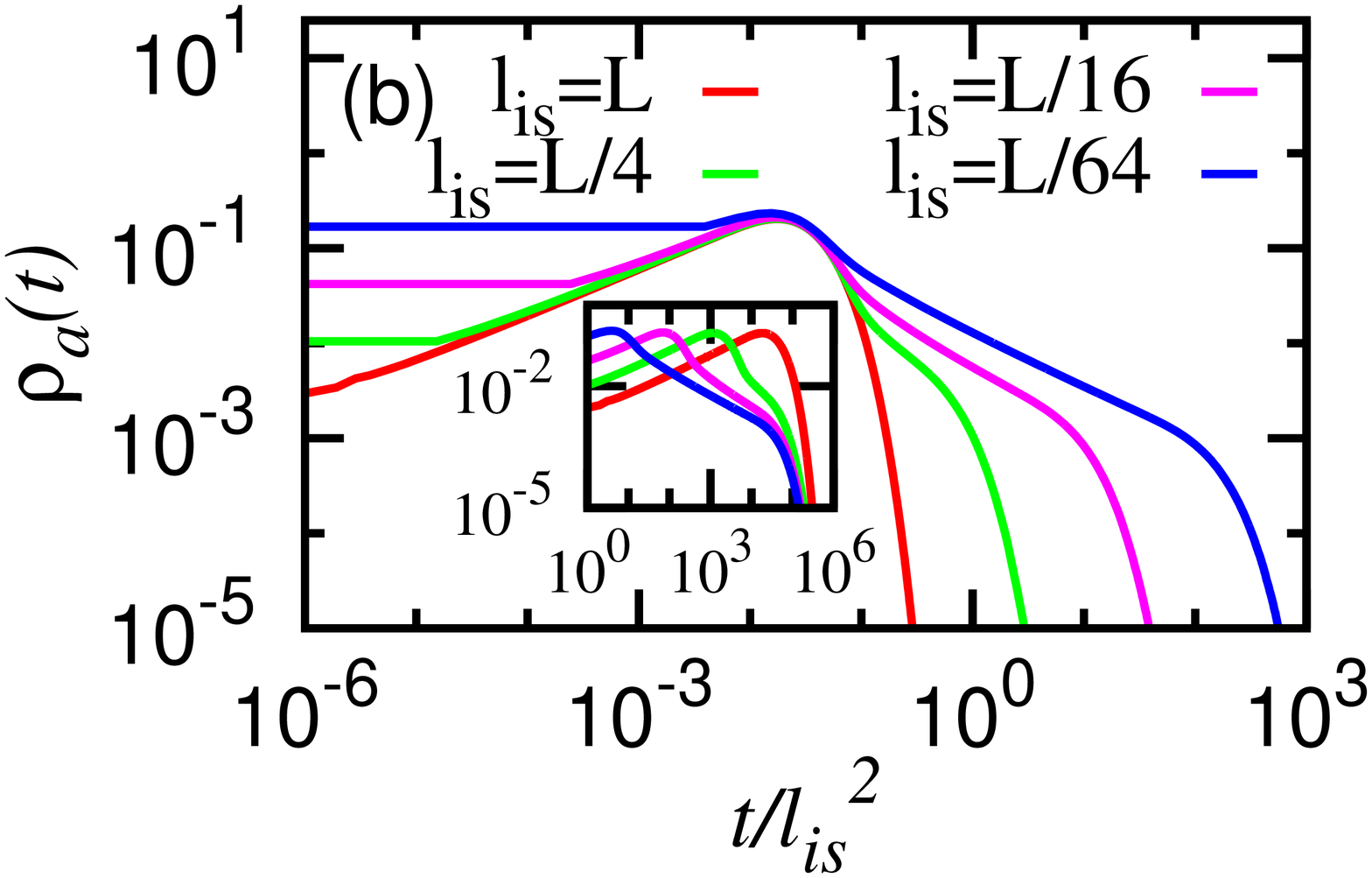}
\vspace{-0.8 cm}
\caption{(Color online) Data collapse for the (a) determination of the time scales  $\tau_{in}$ and $\tau$ for blocked i.c. with block size $L/2$ for different $L$; (b) determination of  $\tau_{in}$ for different block size keeping $L$ fixed ($L=2^{10}$); Insets: Corresponding unscaled data.}
\vspace{-0.2 cm}
\label{fig4}
\end{figure}
The decay part of the dynamics corresponds to $1100\to 1010$ and the diffusion (or spreading) part corresponds to $1101\to 1011$, and $(\langle110\rangle+\langle011\rangle)$ is the total number of active sites. The system starts with two active sites which lie at the ends of a block. Thus initially the process is dominated by diffusion and growth of activity, and this continues until the system becomes uniform. After that the decay process dominates. The cumulative density profile for blocked i.c. with block size $L/2$ is an isosceles triangle with base length $L$, which corresponds to $l_{is}=L$. Therefore, $\tau_{in}$ and $\tau$ are comparable ($\sim L^2$) and hence no power law decay is observed at all (\ref{fig4}(a)). 

Similarly we consider the situations with different block size- $L/2$,~ $L/8,~L/32,$ and $L/128$, filling each block with either $1$ or $0$, keeping $\rho=1/2$ and fixed system size $L=2^{10}$. Filled blocks are placed at equal spatial intervals. Those situations are shown in Fig. \ref{fig4}(b). In both the panels in Fig. \ref{fig4}, the time scales have been rescaled to collapse the data. Thus it is clear from Fig. \ref{fig4} that $\tau_{in}\sim l_{is}^2$ and $\tau\sim L^2$. 

Variation of $\rho_a(t)$ in Fig. \ref{fig4} can be explained in the following way. As we have mentioned that for blocked i.c. initially the dynamics responsible for decay ($1100\to 1010$) of activity is limited to the block ends only and during this period ($\tau_{in}$) the effective dynamics is just diffusion ($1101\to 1011$) which contribute to the growth of activity. Thus, it resembles simple random walk problem and hence the time taken to cover a distance $l$ varies as $l^2$ and the distance covered in time $t$ varies as $t^{1/2}$. Therefore, for both Figs. \ref{fig4}(a) and \ref{fig4}(b), the slip exponent $\Theta=1/2$ and the initial time scale $\tau_{in}$ varies as square of the block size {\it i.e.} $\tau_{in}\sim l_{is}^2$ (in Fig. \ref{fig4}(a) $l_{is}=L$), which are in agreement with those obtained from the numerical simulations.

Now, one can obtain the survival probability exponent $\delta$ form the hyperscaling relation $z(\Theta+\delta)=d-\beta/\nu_\perp$. Here, $d=\beta=\nu_\perp=1$ and this leads to $\delta=-1/2$. With these spreading exponents, the  hyperscaling relation $\Theta=d/z-\alpha-\delta$ results in $\alpha=1/2$, which corresponds to the decay tails in Fig. \ref{fig4}(b), after sufficiently large time. 

Here, it is important to note that the set of spreading exponents $\Theta=1/2$ and $\delta=-1/2$  are obtained for blocked i.c. whereas for natural i.c. we have shown that these exponents are $\Theta=-1/2$ and $\delta=1/2$. Thus, the spreading exponents are also initial condition dependent. However, we have already explained that in case of natural i.c. the initial memory effect is almost absent and the measurements are made in the asymptotic regime,  and hence the universal features are easily extracted from the measurements. Therefore, the set of spreading exponents describing the universal behavior is $\Theta=-1/2$ and $\delta=1/2$.

\begin{wrapfigure}{r}{0.24\textwidth}
\vspace{-0.4 cm}
\hspace{-0.4 cm}
\includegraphics[width=4.4 cm, height=4 cm]{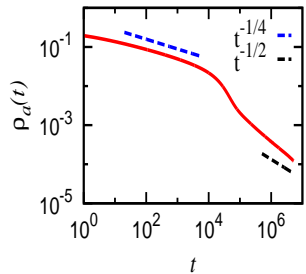}
\vspace{-0.3 cm}
\caption{(Color online) Decay of $\rho_a(t)$ with time using blocked random i.c.}
\vspace{-0.2 cm}
\label{fig5}
\end{wrapfigure}
In order to show the cross over between the two distinct decay behavior, we consider another initial condition where we divide the system into a number of blocks of equal size, then fill one block using random i.c. and then copy the same to other blocks, maintaining the density $\rho=1/2$. With this we can control $l_{is}$ and hence $\tau_{in}$. This makes $\tau_{in}<\tau$ and both the regimes are clearly visible as shown in Fig. \ref{fig5}.

\section{Origin of anomalous behavior and the remedy:}
The origin of the observed unusual behavior is the  memory effect which  gives rise to long range spatial and temporal correlations. There is a time scale $\tau_{in}$ up to which the initial memory persists. The span $\tau_{in}$ depends on the initial condition as well as the dynamics. It takes that much time for the system to built up proper correlations to suppress the microscopic details including initial condition. After that the system reaches the asymptotic regime which captures the universal features.
 
In general there are two processes involved- the initial part is dominated by flattening of the density profile, though the decay process is also coupled to it. Once the system reaches the uniform state the memory of initial information is erased and the  decay process dominates over the diffusion since the background profile has become flat.  The outcome crucially depends on the competition between the two time scales.

There are 3 possibilities- $(i)$ if $\tau_{in}<<\tau$, long range correlations are set up quickly erasing the initial memory, and such systems are free from unusual behaviors and easy to study;  $(ii)$ if $\tau_{in}<\tau$ but long enough, the system may suffer from long lived memory effect which may give rise to undershooting, anomalous scaling,  limitation of system size and computational time etc., and may eventually result in wrong estimate for the critical point and decay exponents.  $(iii)$ if $\tau_{in}\sim\tau$ {\it i.e.} the 
time scales are comparable, there are two possibilities-($a$) if $\alpha_{in}>\alpha_{ss}$,~ $\alpha_{ss}$ dominates and asymptotic regime becomes visible after the initial effect becomes negligible; (b) if $\alpha_{in}<\alpha_{ss}$, the dominating behavior is $\alpha_{in}$ and it will result in a decay exponent which is not the universal one, leading to complete suppression of the universal decay feature. However, the initial decay is not necessarily a power law always. Depending on the dynamics different initial conditions may give rise to different unusual features. 

The ill effects arising from the initial memory can be tactfully eliminated using natural i.c. which ensures $\tau_{in}<<\tau$. Another way is to obtain $\alpha_{ss}$ from the stationary state autocorrelation which is completely free from the initial memory effect. However, stationary state autocorrelation can not be measured accurately if its value is too small in the critical regime.


\section{Conclusion:}
In 1D CLG we have clearly shown that the asymptotic decay exponent $\alpha$ is dependent on the initial condition: $\alpha_{in}=1/4$ for random initial condition  and $\alpha_{in}=1/2$ for natural initial condition. The later is in agreement with that obtained from stationary state autocorrelation, $\alpha_{ss}=1/2$ which is independent of initial condition since it is measured in the stationary state which is unique. The decay exponent $\alpha=1/2$ is consistent with the scaling relations whereas $\alpha=1/4$, obtained from random i.c., shows scaling violation. Thus, natural i.c. and stationary state autocorrelation  capture the universal features of the system whereas random i.c. does not. However, the static exponents do not depend on the initial condition since the stationary state is unique.

Such unusual situation arises because of existence of two competing time scales: (a) $\tau_{in}\sim l_{is}^2$, which is a measure of how long the initial memory persists; (b) $\tau\sim L^2$, which arises from the finite size of the system. For random i.c. $\tau_{in}\sim L^2$ and hence the system encounters finite size effect before reaching the asymptotic regime. Thus for random i.c. asymptotic regime becomes inaccessible.
Natural i.c. ensures $\tau_{in}<<\tau$ which provides easy access to the asymptotic regime and hence captures the universal features.

We expect our results to be also useful to study universal behavior of many other systems where the initial memory may persist for a long time. However, finding  a suitably reactivated (locally perturbed) stationary state namely natural initial condition, may not  always be easy.

~

{\bf Acknowledgement:} We thankfully acknowledge a referee for his/her useful  comments and suggestions.



\begin{thebibliography}{99}
 
\bibitem{Henkel} M. Henkel, H. Hinrichsen, and S. L\"ubeck, {\it Non-Equilibrium phase transitions}, vol. 1 (Springer, Berlin, 2008). 

\bibitem{Marro} G. \'Odor, {\it Universality in nonequilibrium lattice systems} (World Scientific, Singapore, 2008); J. Marro and R. Dickman, {\it Nonequilibrium 
Phase Transitions in Lattice Models} (Cambridge University Press, Cambridge, 1999).

\bibitem{Lubeck} S. L\"ubeck, Int. J. Mod. Phys. B, \textbf{18}, 3977 (2004).
 
\bibitem{dp} H. K. Janssen, Z. Phys. B \textbf{42}, 151 (1981); P. Grassberger, Z. Phys. B \textbf{47}, 365 (1982). 
 
\bibitem{cp} T. E. Harris, Ann. Prob. \textbf{2}, 969 (1974).

\bibitem{epd} D. Mollison, J. R. Statist. Soc. B \textbf{39}, 283 (1977).

\bibitem{Odor} G. \'Odor, Rev. Mod. Phys. \textbf{76}, 663 (2004).

\bibitem{Manna} S. S. Manna, J. Phys. A: Math. Gen. \textbf{24}, L363 (1991).
 
\bibitem{Rossi} M. Rossi, R. Pastor-Satorras, and A. Vespignani, Phys. Rev. Lett. \textbf{85}, 1803 (2000).
 
\bibitem{CLG_Oliveira} M. J. de Oliveira, Phys. Rev. E \textbf{71}, 016112 (2005); U. Basu and P. K. Mohanty, Phys. Rev. E \textbf{79}, 041143 (2009).
 

\bibitem{CLG_Lee}S-G. Lee and S. B. Lee, Phys. Rev. E \textbf{77}, 021113 (2008).
 
\bibitem{CTTP_contracdictory} S. L\"ubeck, Phys. Rev. E \textbf{66}, 046114 (2002); R. Dickman, T. Tom\'e and M. J. de Oliveira, Phys. Rev. E \textbf{66}, 016111 (2002); R. Dickman, Phys. Rev. E \textbf{73}, 036131 (2006). 
 
\bibitem{Maslov}  S. Maslov and Y.-C. Zhang, Physica A \textbf{223}, 1 (1996);
 J. A. Bonachela and M. A. Mu\~noz, Phys. Rev. E \textbf{78}, 041102 (2008).
 
\bibitem{Jensen} I. Jensen and R. Dickman, Phys. Rev. E \textbf{48}, 1710 (1993).

\bibitem{Init} J. F. F. Mendes, R. Dickman, M. Henkel and M. C. Marques, J. Phys A: Math. Gen. \textbf{27}, 3019 (1994).

\bibitem{Tran}  H. K. Janssen, B. Schaub, and B. Schmittmann, Z. Phys. B \textbf{73}, 539 (1989); H. Hinrichsen and G. \'Odor, Phys. Rev. E \textbf{58}, 311 (1998). 

 
\bibitem{Lubeck_split} S. L\"ubeck and P. C. Heger, Phys. Rev. E \textbf{68}, 056102 (2003). 

\bibitem{Wilczek} D. Toussaint and F. Wilczek, J. Chem. Phys. \textbf{78}, 2642 (1983).

 
 \bibitem{fes_dp} M. Basu, U. Basu, S. Bondyopadhyay, P. K. Mohanty, and H. Hinrichsen, Phys. Rev. Lett. \textbf{109}, 015702 (2012). 
 
 \bibitem{Redner} K. Kang and S. Redner, Phys. Rev. Lett. \textbf{52}, 955 (1984);  K. Kang and S. Redner, Phys. Rev. A \textbf{32}, 435 (1985).
 


 
\end{thebibliography}
\end{document}